\documentstyle[12pt]{article}
\def\1{\mbox{I\hspace{-.15em}1}}

\def\b{\begin{equation}}
\def\e{\end{equation}}
\def\bee{\begin{enumerate}}
\def\eee{\end{enumerate}}
\title{Trans-Planckian Scale
\\and\\
Krein Space Quantization}
\author{A. Sojasi$^{1}$\thanks{e-mail:
sojasi@iaurasht.ac.ir}  , M. Mohsenzadeh$^{2}$\thanks{e-mail:
mohsenzadeh@qom-iau.ac.ir}}
\begin{document}

\maketitle {\it \centerline{\it $^{1}$ Department of Physics, Science and Research
Branch, Islamic Azad University, Theran, Iran} \centerline{$^2$
Department of Physics, Qom Branch, Islamic Azad University,  Qom,
Iran}}

\begin{abstract}
In this work, Krein space quantization method is applied to eliminate the
Ultraviolet divergence of Green functions. This paper shows that the power
spectrum of scalar field fluctuations can be calculated in the
limit of short distance physics, {\it trans-Planckian physics}, without
using the usual re-normalization process.

\end{abstract}


\section{Introduction}
It is proven that for the minimally coupled scalar field in de
Sitter space a naturally re-normalized quantum field theory can be
constructed by Krein space quantization method \cite{c1}. It is shown
that in this quantization method, the Hamiltonian operator has a
finite vacuum expectation value. This result is achieved in
different way, without using the normal ordering \cite{c1,c2}. It is
one of the most noticeable results of this quantization method.
Another of the noticeable results of this method is the elimination of the Ultraviolet
divergence of Green functions automatically \cite{c1,c2}. Moreover, since
the minimally coupled scalar field in de Sitter space plays an
important role in the inflationary universe as well as in the
linear quantum gravity \cite{c3}, therefore we refer to previous work in
which the power spectrum of scalar field fluctuations is calculated
by Krein quantization method \cite{c4}. Let us make this remark
that there is a similarity between the Krein quantization method
and usual re-normalization. In the curved space-time, the standard
re-normalization of the Ultraviolet divergence of the vacuum
energy is accomplished by subtracting the local divergence of
Minkowski space \cite{c4,c5} \b
\langle{\Omega}|:T_{\mu\nu}:|{\Omega}\rangle=\langle{\Omega}|T_{\mu\nu}|{\Omega}\rangle-\langle{0}|T_{\mu\nu}|{0}\rangle.\e
In which $ |{\Omega}\rangle $ is the vacuum state in curved space
and $ |{0}\rangle $ is the vacuum state in Minkowski space. The
minus sign in equation (1) can be interpreted as the negative norm
states which is added to the positive norm states \cite{c2}. This
interpretation resembles Krein space quantization method, where the
negative norm states are considered. Indeed in this method field operator is
built by considering two possible solutions of field equations,
positive and negative norm states \cite{c1,c2}. These negative
norm states are defined in Minkowski space and they are not the
solutions of the wave equation in the curved space-time. The auxiliary negative norm
states in our method are similar to the ghost states in the
standard gauge QFT \cite{c6} and they play the role of an
automatic re-normalization tool \cite{c7,c8,c9}. Therefore the auxiliary
negative norm states (ghost states) can neither propagate in the
physical world nor interact with physical states.

 Since the effects of short
distance physics, such as trans-Planckian scale, non
commutative space-time and so on, might be observable in
cosmological scales on the power spectrum of Cosmic Microwave Background
Radiation (CMBR)\cite{c10,c11,c12,c13,c14,c15}, therefore it is reasonable
to present the another ability for Krein space quantization method, which is the
calculation of the power spectrum for scalar field fluctuations in the
early universe. To reach this goal we review the trans-Planckian
scale at first and then we explain how Krein quantization method can
solve our problem.

\section{Trans-Planckian Scale}
Let us consider a background as a de Sitter universe with metric
\cite{c4,c14} \b ds^{2}=dt^{2}-e^{2Ht}{\vec{dx}}^{2}.\e The
equation of motion for scalar field in this background is given by
\cite{c4,c5,c14} \b \ddot{\phi}+3H\dot{\phi}-\nabla^{2}\phi=0. \e
Where field operator can be represented in the form \cite{c5,c6}
\b \phi(\eta,\vec{x})=(2\pi)^{-3/2}\int d^{3}{\vec{k}}
[a_{k}u_{k}(\eta)e^{i\vec{k}.\vec{x}}+a_{k}^{\dag}u_{k}^{*}(\eta)e^{-i\vec{k}.\vec{x}}].\e
And Fourier modes, satisfies the equation \b
u_{k}''+2aHu_{k}-k^{2}u_{k}=0. \e The exact solutions of (5) are \b
u_{k}^{\pm}=\frac{1}{a\sqrt{2k}}(1\mp\frac{i}{k\eta})e^{\mp{ik\eta}}.\e
Continuing in the Heisenberg picture leads to Bogoliubov transformations for $ a
$ and $ a^{\dag} $ between fixed time $ \eta_{0} $ and $ \eta $
\cite{c5,c16} \b a_{k}(\eta)=\alpha
a_{k}(\eta_{0})+\beta^{*}a_{k}^{\dag}(\eta_{0}),\;\; a_{k}^{\dag}(\eta)=\alpha^{*} a_{k}^{\dag}(\eta_{0})+\beta
a_{k}(\eta_{0}).\e

 It is convenient to choose a boundary initial condition by stating that the modes in the limit  $ \eta_{0}\mapsto\infty $ are positive norms of the Bunch-Davis vacuum \cite{c5,c17}. But if the boundary initial condition chosen differently, by the trans-planckian considerations, the vacuum state would not being a Bunch-Davis ones \cite{c10,c11,c12,c13,c14,c15}. Therefore one can choose the vacuum by considering $ a_{k}|{\eta_{0}}\rangle=0 $ in which, $ \eta_{0}=-\frac{\Lambda}{Hk} $ has a finite value and $ \Lambda $ is the energy scale, e.g., Planck scale \cite{c14}. If one assumes $ \frac{\Lambda}{H}>>1 $,  the power spectrum of scalar field fluctuations is given by \cite{c14}
 \b P_{\phi}=(\frac{H}{2\pi})^{2}(1-\frac{H}{\Lambda}\sin(\frac{2\Lambda}{H})). \e
Which is a scale-dependent power spectrum.
\section{Power Spectrum in Krein Space Quantization}
According to the first perspective of \cite{c4} one can choose the
negative norms with respect to the Minkowski background. The
equation of the negative norm states and it's solution is: \b
u_{k}''+\frac{k^{2}}{a^{2}}u_{k}=0 ,\;\;
u_{k,n}=\frac{e^{ik\eta}}{a\sqrt{2k}}.\e The scalar field operator
in this view is given by \cite{c2,c4} $$
\phi(\eta,\vec{x})=(2\pi)^{-3/2}\int d^3{\vec{k}} \{[
u_{k}(\eta)a_{k}e^{-i\vec{k}.\vec{x}}+u_{k}^*(\eta)a_{k}^{\dag}e^{i\vec{k}.\vec{x}}]$$
\b +[
u_{k,n}(\eta)b_{k}e^{-i\vec{k}.\vec{x}}+u_{k,n}^*(\eta)b_{k}^{\dag}e^{i\vec{k}.\vec{x}}]\}.\e
Where $ u_{k} $ and $ u_{k,n} $ are defined respectively in (6) and (9). Also $ a_{k} $ and $ b_{k} $ are two independent operators. Creation and annihilation operators are constrained to obey the following commutation rules \b
[a_{k},a_{k}^{\dag}]=\delta_{kk'},\;\;[b_{k},b_{k}^{\dag}]=-\delta_{kk'}
.\e

Since the nature of fluctuations is statistically Gaussian, then
the variance of it's distribution is given by
\cite{c5,c17,c18,c19} \b
\langle{\phi^{2}}\rangle=\frac{1}{(2\pi)^{3}}\int|u_{k}|^{2}d^{3}\vec{k}-\frac{1}{(2\pi)^{3}}\int|u_{k,n}|^{2}d^{3}\vec{k}.\e
In which the negative norm states satisfy the equation (9) and
positive ones can be written from (6) generally \b
u_{k}=\frac{A}{a\sqrt{2k}}(1-\frac{i}{k\eta})e^{-{ik\eta}}+\frac{B}{a\sqrt{2k}}(1+\frac{i}{k\eta})e^{{ik\eta}}.\e
which by considering initial condition with finite $ \eta_{0} $
and normalization condition for $ A $ and $ B $, e.g., $
|A|^{2}-|B|^{2}=1 $, we have \cite{c14} \b
B=A\frac{ie^{-2ik\eta_{0}}}{2k\eta_{0}+i},\;\;
|A|^{2}=\frac{1}{1-|\gamma|^{2}},\;\;
\gamma=\frac{i}{2k\eta_{0}+i}.\e

The next step is calculation of
variance of distribution with using Krein
quantization method. Using (9) and (13), we may write (12) as $$
\langle{\phi^{2}}\rangle=\frac{1}{(2\pi)^{3}}\int\frac{d^{3}\vec{k}}{2ka^{2}}\{[(|A|^{2}+|B|^{2})(1+\frac{1}{(k\eta)^{2}})+A^{*}Be^{2ik\eta}(1+\frac{i}{k\eta})^{2}$$ \b +B^{*}Ae^{-2ik\eta}(1-\frac{i}{k\eta})^{2}]
\}-\frac{1}{(2\pi)^{3}}\int\frac{d^{3}\vec{k}}{2ka^{2}}.\e With
substituting (14) in (15) one can find $$
\langle{\phi^{2}}\rangle=\frac{1}{(2\pi)^{3}}\int\frac{d^{3}\vec{k}}{2ka^{2}}\{|A|^{2}[(1+\frac{1}{1+4k^{2}\eta_{0}^{2}})(1+\frac{1}{(k\eta)^{2}})+\gamma
e^{2ik(\eta-\eta_{0})}(1+\frac{i}{k\eta})^{2}$$ \b +\gamma^{*}
e^{-2ik(\eta-\eta_{0})}(1-\frac{i}{k\eta})^{2}]\}
-\frac{1}{(2\pi)^{3}}\int\frac{d^{3}\vec{k}}{2ka^{2}}.\e
Calculating (16) in the limit $ \eta\mapsto 0 $   , $
\frac{\Lambda}{H}>>1 $ , and substitute $ \eta=-\frac{1}{aH} $
leads to \b \langle{\phi^{2}}\rangle=\frac{1}{(2\pi)^{3}}\int
d^{3}\vec{k}[(\frac{H^{2}}{2k^{3}}+\frac{1}{2ka^{2}})-\frac{H^{3}}{2\Lambda
k^{3}}\sin(\frac{2\Lambda}{H})]-\frac{1}{(2\pi)^{3}}\int\frac{d^{3}\vec{k}}{2ka^{2}}.\e
We can see explicitly vanishing Ultraviolet divergence from (17).

\section{Conclusions}
In this work we shows that Krein space quantization method can be
convenient in constructing naturally re-normalized QFT. We
investigated that it is possible to remove Ultraviolet divergence of
Wightman two-point function which is applicable in inflationary
cosmology, by considering negative norm states.

\noindent {\bf{Acknowlegements}}: We would like to thank M. V.
Takook for his valuable help.

\end{document}